\documentclass[pra,aps,twocolumn,floatfix,showpacs,bibnotes]{revtex4}
\usepackage{epsfig}

\begin{document}

\title{Quantum inference of states and processes}

\author{Miroslav Je{\v{z}}ek}
\email{jezek@optics.upol.cz}

\author{Jarom\'{\i}r Fiur\'{a}\v{s}ek}

\author{Zden\v{e}k Hradil}

\affiliation{Department of Optics, Palack\'{y} University,
             17. listopadu 50, 77200 Olomouc, Czech Republic}

\date{\today}

\begin{abstract}
The maximum-likelihood principle unifies inference of quantum states
and processes from experimental noisy data. 
Particularly, a generic quantum process may be estimated simultaneously 
with unknown quantum probe states provided that measurements on probe
and transformed probe states are available.
Drawbacks of various approximate treatments are considered.
\end{abstract}

\maketitle

\section{Introduction}

The various quantum-state reconstruction techniques developed
during recent years have made it possible to completely reconstruct
an unknown state of a quantum mechanical system provided that many
identical copies of the state are available. These reconstruction
methods are nowadays routinely applied to the evaluation of the
experiments where quantum states are generated, manipulated and
transmitted. The field was pioneered in the beginning of nineties
in quantum optics, where the optical homodyne tomography has been
devised for reconstruction of the quantum state of traveling light
field \cite{Vogel_Risken_1989,Smithey_Beck_Raymer_Faridani_1993,%
D'Ariano_Leonhardt_Paul_1995,Schiller_Breitenbach_et_al_1996,%
Breitenbach_Schiller_Mlynek_1997}.
Since then, many other reconstruction methods applicable to various
physical systems have been developed
\cite{Raymer_Beck_McAlister_1994,Jones_1994,%
D'Ariano_Macchiavello_Paris_1994,Dunn_Walmsley_Mukamel_1995,%
Leonhardt_Paul_D'Ariano_1995,Leonhardt_Raymer_1996,%
Banaszek_Wodkiewicz_1996,Leibfried_et_al_1996,%
Kurtsiefer_Pfau_Mlynek_1997,Lutterbach_Davidovich_1997,%
Opatrny_Welsch_1997,Opatrny_Welsch_Vogel_1997,%
Bodendorf_et_al_1998}.
The inference of quantum states plays very important role
in the present-day experiments
\cite{Lvovsky_Hansen_et_al_2001,Lvovsky_Babichev_2002,
Opatrny_Korolkova_Leuchs_2002,Burlakov_et_al_2002}.

Most of the reconstruction methods, such as the direct sampling in
optical homodyne tomography, are based on a direct linear inversion of
the experimental data. This approach is conceptually simple and feasible.
However, it may lead to certain unphysical artifacts such as the negative
eigenvalues of the reconstructed density matrix.
In order to avoid these unphysical artifacts, an estimation method
based on statistical maximum-likelihood principle
has been devised for the reconstruction of a generic quantum state
\cite{Hradil_1997,Hradil_Summhammer_Rauch_1999,%
Hradil_Summhammer_Badurek_Rauch_2000,Hradil_Summhammer_2000,%
Banaszek_D'Ariano_Paris_Sacchi_2000,Rehacek_Hradil_Jezek_2001}.
This approach guarantees the positive semidefiniteness and trace
normalization of the reconstructed density matrix.
These necessary conditions are incorporated as constraints,
so as a certain prior information from the statistical point of view.
Remarkably, the maximum likelihood estimation can be interpreted as
a genuine generalized quantum measurement
\cite{Hradil_Summhammer_Rauch_1999,Hradil_Summhammer_2000}
and can be related to the information gained by optimal measurement
and the Fisher information \cite{Rehacek_Hradil_2002_INF}.

Given current interest in the quantum-information processing, it is
of paramount importance to reconstruct not only the quantum states but also
the transformations of these states---the quantum mechanical processes.
The examination of quantum communication channels and the evaluation of
the performance of quantum gates are the examples of practical applicability
of quantum-process reconstruction
\cite{Poyatos_Cirac_Zoller_1997,Chuang_Nielsen_1997,%
D'Ariano_Maccone_1998,Luis_Sanchez-Soto_1999,%
Gutzeit_Wallentowitz_Vogel_2000,Childs_Chuang_Leung_2001,%
Fischer_Mack_Cirone_Freyberger_2001,D'Ariano_LoPresti_2001}.
All necessary properties of the deterministic quantum transformations,
namely the complete positivity and trace preservation
can be again incorporated  within the maximum-likelihood approach
as the appropriate constraints \cite{Fiurasek_Hradil_2001}.
Compared with other reconstruction methods the maximum-likelihood
approach seems to be computationally more difficult.
Therefore several  simplifications and approximations of
the maximum-likelihood technique have been suggested recently
\cite{James_Kwiat_Munro_White_2001,Sacchi_2001_A}.

In this paper we present a unified approach to the maximum-likelihood
reconstruction of quantum states and quantum processes.
Extremal equations for the reconstructed  quantum state and for
quantum process are derived in Section II.  These equations can easily be
solved numerically by means of repeated iterations.
Particular attention will be  paid to the probing of the quantum process
by entangled states which attracted considerable attention recently.
In Section III we consider a realistic scenario where an unknown quantum
transformation is probed by unknown states and the measurements are
performed on both the input and output states. We propose a method for
simultaneous estimation of the unknown probe states and the quantum
process from the collected experimental data.
The comparison of the exact maximum-likelihood method with
the approximate ones is carried out in Section \ref{comparison}.
Finally, the conclusions are given in Section V.

\section{Reconstruction of quantum process}
\label{process}


Let us start with a brief review of the maximum-likelihood
reconstruction of a  quantum state.
We assume a finite number $N$ of identical samples
of the physical system, each in the same but unknown quantum state
described by the density operator $\rho$. Having
these systems our task is to infer the unknown quantum state $\rho$
from the results of the measurements performed on them. We consider
the positive operator-valued measure (POVM) \cite{Helstrom_QDET}
$\Pi_l$ that yields probabilities $p_l$ of individual outcomes,
\begin{equation}  \label{state_probabilities}
  p_l = {\rm Tr} \left[ \rho \Pi_l \right], \qquad
  p_l \geq 0, \qquad
  \sum_l p_l = 1.
\end{equation}
If the POVM $\Pi_l$ is tomographically complete it is possible
to determine the true state $\rho$ directly by inverting the linear
relations (\ref{state_probabilities}) between the probabilities $p_l$
and the elements of the density matrix $\rho$.
However, there is no way how to find out the exact probabilities $p_l$
since only a finite number $N$ of samples of physical systems
can be investigated. In the case of $N_l$ occurrences
of outcomes $\Pi_l$ the relative detection frequencies $f_l = N_l/N$
represent the only data that could be used for reconstructing
the true state $\rho$. The maximum-likelihood approach to this
reconstruction problem consists in finding a density operator
$\rho_{\rm est}$ that generates through Eq.~(\ref{state_probabilities})
probabilities $p_l$ which are as close to the observed frequencies $f_l$
as possible \cite{Hradil_1997,Rehacek_Hradil_Jezek_2001},
\begin{eqnarray}
  &
  {\displaystyle
  \rho_{\rm est} = \arg\max_{\rho} {\cal L}[f_l,p_l(\rho)], }
  &  \label{state_estimation}  \\
  &
  {\displaystyle
  {\cal L}[f_l,p_l(\rho)] = \sum_l f_l \ln p_l. }
  &  \label{state_likelihood}
\end{eqnarray}
The measure ${\cal L}[f_l,p_l(\rho)]$ of the distance between the
probability distribution $p_l$ and the detected relative frequencies
$f_l$ seems to be arbitrary. However, it can be shown that
the reconstruction procedure can be interpreted as a generalized
POVM measurement if the log-likelihood measure (\ref{state_likelihood})
is used \cite{Hradil_Summhammer_Rauch_1999,Hradil_Summhammer_2000}.
The maximum-likelihood principle has been successfully applied
to many problems of quantum-information processing, for example
to reconstruction of the spin state of an electron or polarization
state of a photon \cite{Hradil_Summhammer_Badurek_Rauch_2000},
reconstruction of entangled spin state
\cite{Rehacek_Hradil_Jezek_2001}, estimation of quantum measurement
\cite{Fiurasek_2001_POVM}, design of the optimal discrimination
device for communication through a noisy quantum channel
\cite{Jezek_2002} and characterization of the universal cloning
machine \cite{Sacchi_2001_B}.

The challenging problem of the maximization (\ref{state_estimation})
of the log-likelihood functional (\ref{state_likelihood}) on the space
of positive semidefinite operators $\rho$, ${\rm Tr}[\rho]=1$, has been
treated with the help of the numerical up-hill simplex method
\cite{Banaszek_D'Ariano_Paris_Sacchi_2000}. A more analytical approach
to the problem involves a formulation of nonlinear extremal operator
equation for the density matrix that maximizes the log-likelihood
functional \cite{Hradil_1997,Hradil_Summhammer_Rauch_1999,%
Rehacek_Hradil_Jezek_2001},
\begin{equation}  \label{state_extremal_eq}
  \rho = \mu^{-1} R \rho, \quad
  R = \sum_l \frac{f_l}{p_l} \Pi_l,
\end{equation}
where the Lagrange multiplier $\mu$  reads
\begin{equation}  \label{state_extremal_eq_mu}
  \mu = {\rm Tr}[R\rho] = \sum_l f_l=1.
\end{equation}
The crucial advantage of the equation (\ref{state_extremal_eq})
is that it is suitable for iterative solution, as has been
demonstrated on many particular reconstruction problems. A combination
of equation (\ref{state_extremal_eq}) and hermitian conjugate equation
leads to the symmetric extremal equations in the manifestly positive
semidefinite form \cite{Fiurasek_2001_POVM},
\begin{equation}  \label{state_extremal_sym_eq}
\rho = \mu^{-2} R \rho R, \qquad
\mu = \left( {\rm Tr}[R \rho R] \right)^{1/2}.
\end{equation}
The iterations
\begin{equation}  \label{state_extremal_sym_eq_iter}
  \rho^{(n+1)} = {\mu^{(n)}}^{-2} R^{(n)} \rho^{(n)} R^{(n)}
\end{equation}
preserve the positive semidefiniteness and trace normalization
of the density operator $\rho$.


While density operator describes the state of physical system,
the linear completely positive (CP) map describes the generic
transformation of physical system from quantum state $\rho_{\rm in}$
to quantum state $\rho_{\rm out}$.
The mathematical formulation of CP maps relies on the isomorphism
between linear CP maps ${\cal M}_{S}$ from operators on the Hilbert space
${\cal H}$ to operators on the Hilbert space ${\cal K}$ and positive
semidefinite operators ${S}$ on Hilbert space ${\cal H}\otimes{\cal K}$
\cite{Hellwig_Kraus_1970,Jamiolkowski_1972,Schumacher_1996},
\begin{equation}  \label{CP_map}
  \rho_{\rm out} =
  {\cal M}_{{S}} \left[ \rho_{\rm in} \right] =
  {\rm Tr}_{\cal H} \left[
  {S} \, \rho_{\rm in}^{\rm T} \! \otimes \openone_{\cal K}
  \right],
\end{equation}
where $\openone_{\cal K}$ is an identity operator on the space ${\cal K}$
and ${\rm T}$ denotes the transposition.
The deterministic quantum transformations preserve the trace of the
transformed operators, ${\rm Tr}_{\cal K}[\rho_{\rm out}]={\rm
Tr}_{\cal{H}}[\rho_{\rm in}]$. Since this must hold for any $\rho_{\rm in}$
the operator $S$ must satisfy the condition
\begin{equation}  \label{tr_preserving}
 {\rm Tr}_{\cal K} [{S}] = \openone_{\cal H},
\end{equation}
where $\openone_{\cal H}$ is an identity operator on space ${\cal H}$.
The condition (\ref{tr_preserving}) effectively represents
$({\rm dim}{\cal H})^2$ real constraints.

Making use of the formalism (\ref{CP_map}) we may formulate the exact
maximum-likelihood principle for estimated CP map ${S}$ in
a particularly simple and transparent form and we can also
straightforwardly  extend the results obtained in Ref.
\cite{Fiurasek_Hradil_2001} to the cases when the input and output
Hilbert spaces have different dimensions.

Let $\rho_m$ denote the various input states from the space ${\cal H}$
that are used for the determination of the quantum process.
Measurements described by POVMs $\Pi_{ml}$ are carried out on each
corresponding output state from space ${\cal K}$. Let $f_{ml}$
denote the relative frequency of detection of the POVM element
${\Pi}_{ml}$. The estimated operator ${S}$ should maximize the
constrained log-likelihood functional
\cite{Fiurasek_Hradil_2001,Sacchi_2001_A}
\begin{eqnarray}
  &
  {\displaystyle {\cal L}_{\rm c}[f_{ml},p_{ml}({S})] =
  \sum_{m,l} f_{ml} \ln p_{ml} - {\rm Tr}[\Lambda {S}], }
  &  \label{process_likelihood}  \\
  &
  {\displaystyle
  p_{ml} = {\rm Tr}\left[ {S} \, \rho_m^{\rm T} \otimes \Pi_{ml} \right], }
  &  \label{process_probabilities}
\end{eqnarray}
where  ${\Lambda} = \lambda \otimes \openone_{\cal K}$ and $\lambda$ is
the matrix of Lagrange multipliers that account for the
trace-preservation condition (\ref{tr_preserving}).
The extremal equations for ${S}$ can be obtained by varying functional
(\ref{process_likelihood}) with respect to ${S}$, which leads to
\begin{equation}  \label{process_extremal_eq}
  {S} = {\Lambda}^{-1} {K} {S}, \qquad
  {K} = \sum_{m,l} \frac{f_{ml}}{p_{ml}} \rho_m^{\rm T} \otimes \Pi_{ml}.
\end{equation}
Further we have from Eq. (\ref{process_extremal_eq}) that
${S} = {S} {K} {\Lambda}^{-1}$. When we insert this expression in the
right-hand side of Eq. (\ref{process_extremal_eq}), we finally arrive at
symmetrical expression suitable for iterations,
\begin{equation}
\label{process_extremal_sym_eq}
  {S} = {\Lambda}^{-1} {K} {S} {K} {\Lambda}^{-1}.
\end{equation}
The Lagrange multiplier ${\lambda}$ must be determined from the constraint
(\ref{tr_preserving}). On tracing Eq. (\ref{process_extremal_sym_eq}) over
space ${\cal K}$ we obtain quadratic equation for ${\lambda}$
which may be solved as
\begin{equation}  \label{process_extremal_sym_eq_lambda}
  \lambda = \left( {\rm Tr}_{\cal K}[{K} {S} {K}] \right)^{1/2}.
\end{equation}
The operator $\Lambda$ is positive definite because ${K} {S} {K}$
is positive definite operator.
The system of coupled Eqs. (\ref{process_extremal_sym_eq}) and
(\ref{process_extremal_sym_eq_lambda})
may be conveniently  solved numerically by means of repeated iterations,
starting from some unbiased CP map, for example
${S}^{(0)} = \openone_{{\cal H} \otimes {\cal K}} / ({\rm dim}{\cal K})$.
It is important to note that Eq.~(\ref{process_extremal_sym_eq})
preserves the positive semidefiniteness of ${S}$ and also the constraint
${\rm Tr}_{\cal K}[{S}] = \openone_{\cal H}$ is satisfied at each
iteration step.

The density matrix  $S$  representing the
CP map ${\cal{M}}_S$ can be in fact prepared physically in the laboratory
if we first prepare a maximally entangled state on the Hilbert space
${\cal{H}}\otimes \cal{H}$
and then apply a CP map to one part of this entangled state. In this way
the quantum-process tomography can be transformed to the quantum-state
tomography. More generally, this suggests that it may be useful to employ
entangled quantum states as probes of the unknown quantum process
\cite{D'Ariano_LoPresti_2001}.

Let $\rho_{m,AB}$ denote the entangled state on the Hilbert space
${\cal{H}}_A\otimes {\cal{H}}_B$ that serves as a probe of the CP map
$S$ that is applied to the subsystem $A$. A joint generalized
measurement described by the POVMs $\Pi_{ml}$ if performed on the output
Hilbert space ${\cal{K}}\otimes{\cal{H}}_B$. The log-likelihood
functional has the form (\ref{process_likelihood}),
only the formula for the probability $p_{ml}$ changes to
\begin{equation}
p_{ml}= {\rm Tr}_{{\cal{H}}_A{\cal{H}}_B\cal{K}}
[(S \otimes \openone_{{\cal{H}}_B})
(\rho_{m,AB}^{T_A}\otimes \openone_{\cal{K}})
(\openone_{{\cal{H}}_A}\otimes \Pi_{ml})],
\end{equation}
where $T_A$ stands for the partial transposition in the subsystem $A$.
Consequently, the operator $K$ appearing in the extremal Eqs.
(\ref{process_extremal_sym_eq}) and  (\ref{process_extremal_sym_eq_lambda})
must be calculated as follows,
\begin{equation}
K=\sum_{m,l}\frac{f_{ml}}{p_{ml}}
{\rm Tr}_{{\cal{H}}_B}
[(\rho_{m,AB}^{T_A}\otimes \openone_{\cal{K}})
(\openone_{{\cal{H}}_A}\otimes \Pi_{ml})].
\end{equation}
Apart from these modifications of $p_{ml}$ and $K$ one can proceed as
before and solve Eqs.~(\ref{process_extremal_sym_eq}) and
(\ref{process_extremal_sym_eq_lambda}) by means of repeated iterations.

\section{Quantum process measurement by unknown probe quantum states}

\label{simultaneous}


Up to now  quantum states and processes have been treated independently. 
However, this is just a simplification typical for the realm of physical 
experiments. Widely accepted strategy how to approach a complex problem
is to specify some partial subproblems, address them separately and
merge the solutions. This technique  usually gives good answer in the
technical sense.
Though this is possible even in quantum theory, there are no
fundamental reasons for such a factorization. To consider
the full problem without splitting it into isolated subproblems is
technically more advanced but could be advantageous. This strategy
will be demonstrated on the synthesis of the problems treated separately
in the previous section. Let us assume the estimation of the generic
process with the help of set of probe states, identity of which is also
unknown. What is only known to the experimentalists are the output of
certain measurements performed on the ensemble of probe states and on the
ensemble of transformed probe states. In this sense all the considerations
are done {\em ab initio\/}, since only  results of generic measurements
are required. A quantum object could be considered as known only to the
extent specified by some preceding measurements. All the physically
relevant results will be derived exclusively from the acquired data,
where input states and their transformation are inseparably involved.
States and their transformation should be considered as quantum
objects. As such they are affected by quantum fluctuations, since in
every experiment a certain portion of the noise will be present on 
the microscopic level.

\begin{figure}[bt]
  \centerline{\epsfig{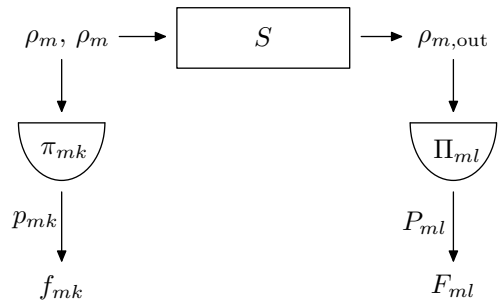}}
  \caption{Scheme of setup for the generalized
           measurement of quantum process using unknown
           quantum states as probes.}
  \label{fig_simultaneous}
\end{figure}

In the following the probe quantum states $\rho_m$ will be treated as
unknown mixed states and they will be inferred together with the unknown
quantum process ${S}$. In accordance with the theory presented above let
us consider the set of probe  states $\rho_m$ on the space ${\cal H}$.
By means of unknown quantum process ${S}$ these states are transformed
onto output states $\rho_{m,{\rm out}}$ in the  space ${\cal K}$.
The observation must be more complex now involving the detection on the
ensemble of both the input and the output states. For this purpose
the corresponding POVM elements will be denoted by $\pi_{mk}$ and
$\Pi_{ml}$. The diagram involving detected signals and measurements is
shown in Fig.~\ref{fig_simultaneous}.
Let $f_{mk}$ denotes the relative frequency of detection of the POVM
element $\pi_{mk}$ in the input space ${\cal H}$ and $F_{ml}$ denotes
the relative frequency of detection of the POVM element $\Pi_{ml}$
in the output space ${\cal K}$. The frequencies $f_{mk}$,
$\sum_k f_{mk} = 1$, and $F_{ml}$, $\sum_l F_{ml} = 1$,
approximate the true probabilities $p_{mk}$ and $P_{ml}$ of individual
outcomes, respectively,
\begin{equation}  \label{simultaneous_probabilities}
  \begin{array}{c}
  {\displaystyle
    p_{mk} =
      {\rm Tr}_{\cal H}\!\left[ \rho_m \pi_{ml} \right], } \\
  ~ \vspace*{-2mm} ~ \\
  {\displaystyle
    P_{ml} =
      {\rm Tr}_{\cal K}\!
      \left[ \rho_{m,{\rm out}} \Pi_{ml} \right] =
      {\rm Tr}\!
      \left[ {S} (\rho_m^{\rm T} \otimes \Pi_{ml}) \right], }
  \end{array}
\end{equation}
where the relation (\ref{CP_map}) was used. The estimated process ${S}$
and probe states $\rho_m$ should maximize the constrained log-likelihood
functional
\begin{equation}  \label{simultaneous_likelihood}
  \begin{array}{c}
    {\displaystyle {\cal L}_{\rm c} =
    \sum_{m,k} f_{mk} \ln{ p_{mk} } +
    \sum_{m,l} F_{ml} \ln{ P_{ml} } - } \\
    ~ \vspace*{-2mm} ~ \\
    {\displaystyle
    - \sum_m \mu_m {\rm Tr}\left[ \rho_m \right]
    - {\rm Tr}\left[ \Lambda {S} \right]. }
  \end{array}
\end{equation}
The additivity of log likelihood reflects the independence of
observations performed on the input and output states with the
same degree of credibility. The Lagrange multipliers $\mu_m$ and
$\Lambda = \lambda \otimes \openone_{\cal K}$ fix necessary
constraints---the trace normalization of the states,
${\rm Tr}[\rho_m] = 1$, and the trace-preserving property
(\ref{tr_preserving}) of the process ${S}$.

The coupled extremal equations for the probe states $\rho_m$ and
for the process ${S}$ can be obtained by varying
(\ref{simultaneous_likelihood}) with respect to independent variables
$\rho_m$ and ${S}$, which yields
\begin{eqnarray}
  &
  {\displaystyle
  \mu_m^{-2} \, {R}_m \rho_m {R}_m = \rho_m,  } &
  \label{simultaneous_extremal_sym_eq__states} \\
  &
  {\displaystyle
  \Lambda^{-1}  {K} {S} {K}  \Lambda^{-1} = {S},  } &
  \label{simultaneous_extremal_sym_eq__process}
\end{eqnarray}
where
\begin{eqnarray}
  &
  \begin{array}{c}
  {\displaystyle {R}_m =
    \sum_k \frac{f_{mk}}{p_{mk}} \pi_{mk} +
  } \\
  +
  {\displaystyle
    {\rm Tr}_{\cal{K}}\left[ {T}_{\cal H} {S}
    \left( \openone_{\cal H} \otimes
      \sum_l \frac{F_{ml}}{P_{ml}} \Pi_{ml}
    \right)
    \right],
  }
  \end{array}
  &
  \label{simultaneous_extremal_sym_eq__states_kernel} \\
  &
  {\displaystyle
    {K} =
      \sum_{m,l} \frac{F_{ml}}{P_{ml}}
      \rho_m^{\rm T} \otimes \Pi_{ml},
  }
  &
  \label{simultaneous_extremal_sym_eq__process_kernel}
\end{eqnarray}
and ${T}_{\cal H}$ is operator of partial transposition
in space ${\cal H}$ acting on space ${\cal H} \otimes {\cal K}$.
The Lagrange multipliers can be determined from
the appropriate constraints,
\begin{eqnarray}
  &
  {\displaystyle
  \mu_m = \left( {\rm Tr}_{\cal H}\!\left[
  {R}_m \rho_m {R}_m
  \right] \right)^{\frac{1}{2}}, }
  &
  \label{simultaneous_extremal_sym_eq_mu} \\
  &
  {\displaystyle
  {\lambda} = \left( {\rm Tr}_{\cal K}\!\left[
  {K} {S} {K}
  \right] \right)^{\frac{1}{2}}.
  } &
  \label{simultaneous_extremal_sym_eq_lambda}
\end{eqnarray}
All necessary properties of the quantum states $\rho_m$
and the quantum process ${S}$ are satisfied during
the iterative solution of the extremal equations
(\ref{simultaneous_extremal_sym_eq__states})--%
(\ref{simultaneous_extremal_sym_eq_lambda}).

\begin{figure}[hbt]
  \centerline{\epsfig{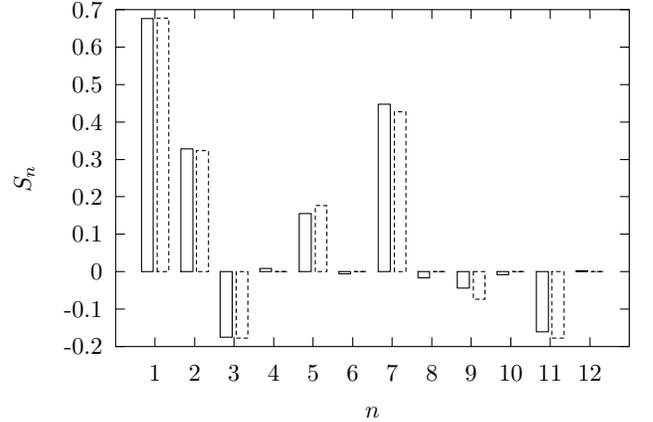}}
  \caption{Elements of the reconstructed quantum process
           (solid) are compared with the theoretical
           ones (dashed) for the rotating-damping channel
           and $20$ various probe states.}
  \label{fig_process}
\end{figure}

In the rest of this section we illustrate the developed
method on the estimation of a quantum process ${S}$ that
transforms one qubit state to another one,
${\rm dim}{\cal H} = {\rm dim}{\cal K} = 2$.
The process ${S}$ under consideration consists of a unitary
$\pi/4$-rotation in $xz$-plane of the Bloch space and
a subsequent non-unitary damping. The unitary part of the
process can be represented by its action on ortogonal
states $|0\rangle$ and $|1\rangle$,
\begin{equation}  \label{pi4_rotation_process}
  \begin{array}{l}
    |0\rangle \, \rightarrow \, \cos\theta \, |0\rangle +
                                \sin\theta \, |1\rangle, \\
    |1\rangle \, \rightarrow \, \cos\theta \, |1\rangle -
                                \sin\theta \, |0\rangle,
  \end{array}
\end{equation}
where $\theta = \pi/8$. The non-unitary part of the process
is described by the operator $d\,{D} + (1-d)\,{E}$,
where we chose $d = 1/2$. The process
${D} = \openone_{{\cal H}\otimes{\cal K}}/2$ is totally
depolarizing channel that maps all states to the maximally
mixed state and ${E}$ is the identity transformation.
We have performed numerical simulations of the $\pi_{mk}$
and $\Pi_{ml}$ measurements for $M=20$ input probe states
$\rho_m$ and the corresponding transformed states
$\rho_{m,{\rm out}}$ respectively. The mixed states $\rho_m$
have been randomly generated. We consider a convenient
experimental realization where the same measurements are
performed on all input as well as output states. In the present
example this POVM measurement consists of tomographically
complete set of projective measurements in $x$, $y$ and $z$
directions, each made on $N=1000$ identical samples of the probe
states $\rho_m$ before and after transformation. Therefore, the
total number $6MN$ of the probe states have been used up.
Theoretical probabilities $p_{mk}$ and $P_{ml}$ have been
evaluated according to Eq.~(\ref{simultaneous_probabilities}).
They represent mean values of the multinomial distributions of
the relative frequencies $f_{mk}$ and $F_{ml}$. Corresponding
variances are approximately given by $p_{mk}(1-p_{mk})/N$ and
$P_{ml}(1-P_{ml})/N$, respectively. The data $f_{mk}$ and
$F_{ml}$ have been obtained by means of Monte-Carlo simulation.
Subsequently, we have iteratively solved extremal
equations (\ref{simultaneous_extremal_sym_eq__states})--%
(\ref{simultaneous_extremal_sym_eq_lambda}).
Result of this procedure is shown in Fig.~\ref{fig_process}.
Only $12$ real independent elements of estimated process are
plotted in form of a vector $\{{S}_n\}_{n=1}^{12}$.
The estimated values are well corresponding to the true ones.

\begin{figure}[th]
  \centerline{\epsfig{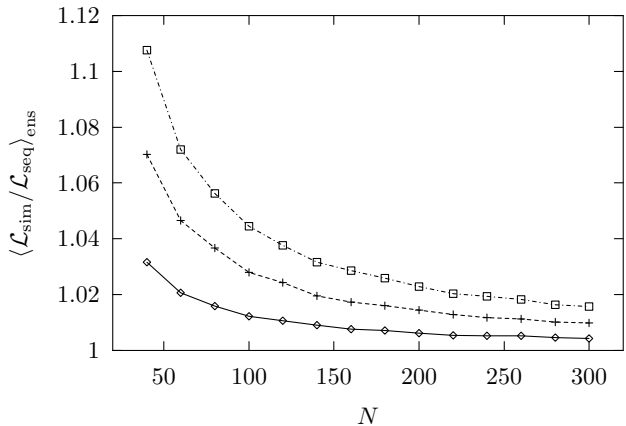}}
  \caption{The average ratio of the likelihood attained by simultaneous
           reconstruction of a quantum process and probe states
           and the likelihood attained by sequential one.
           The process is probed by $15$ (diamond),
           $30$ (plus) and $45$ (square) quantum states.
           The ratio decreases with the increasing number $N$
           of measurements.}
  \label{fig_simseq}
\end{figure}

The simultaneous reconstruction discussed above yields
a higher likelihood of estimated quantum objects than
separate reconstructions of probe states and
a quantum process. This seems to be a general rule.
The likelihood ${\cal L}_{\rm sim}$ obtained by simultaneous
reconstruction (\ref{simultaneous_extremal_sym_eq__states})--%
(\ref{simultaneous_extremal_sym_eq_lambda}) of the quantum process
${S}$ and the probe states $\rho_m$ is always higher than the sum
${\cal L}_{\rm seq}$ of likelihoods ${\cal L}_{\rho_m}$ obtained
by the separate reconstructions (\ref{state_extremal_sym_eq}) of
the probe states $\rho_m$ from data $f_{mk}$, $F_{ml}$ and
likelihood ${\cal L}_{S}$ of the estimated quantum process
${S}$ (\ref{process_extremal_sym_eq})--%
(\ref{process_extremal_sym_eq_lambda}),
where the reconstructed probe states are utilized. 
The ratio ${\cal L}_{\rm sim}/{\cal L}_{\rm seq}$
averaged over an ensemble of possible experimental data
is plotted in Fig.~\ref{fig_simseq} for several numbers
of probe states and various numbers $N$ of measurements.
The true process ${S}$ and the POVM measurements $\pi_{mk}$,
$\Pi_{ml}$ are the same as in the previous example.
A significant improvement is obtained by using the proposed
simultaneous reconstruction method in the case of small number
$N$ of measurements, so in the case of noisy data.
The quantitative difference between simultaneous and sequential
reconstruction procedures changes to qualitative one for a
tomographically incomplete POVM measurement in the input or
output space. Data acquired by such a measurement could be insufficient
for the sequential reconstructions, however, they can be sufficient
for the simultaneous one. For example, projective measurements
in $x$, $y$ directions in the input space and projective
measurements in $y$, $z$ directions in the output space
represent this case. Thus the presented simultaneous reconstruction
technique is applicable to the problems, where routine sequential
methods fail.

\section{Approximate methods}

\label{comparison}


Recently, approximate reconstruction methods based on the maximum
likelihood have been presented. Two ways can be followed to modify
the exact maximum-likelihood principle---either simplification
of the distance measure (\ref{state_likelihood})
\cite{James_Kwiat_Munro_White_2001} or releasing some constraints
on quantum states and processes \cite{Sacchi_2001_A}.

For large number $N$ of identical samples of quantum states available
for inspection going before the state reconstruction the relative
frequencies $f_l$ fluctuate around the true values $p_l$ according
to the multidimensional Gaussian distribution that approximates the
exact multinomial one,
\begin{equation}  \label{multinomial_to_Gauss}
  \begin{array}{cc}
    {\displaystyle
    \prod_l p_l^{f_l}
    \rightarrow
    \prod_l \exp \left[ -\frac{(f_l - p_l)^2}{2 \sigma_l^2} \right],} \\
    \vspace*{-2mm}\\
    {\displaystyle
    \sigma_l^2 \approx p_l (1 - p_l) / N.}
  \end{array}
\end{equation}
Accordingly, the exact likelihood functional (\ref{state_likelihood})
can be replaced by the approximate one
\cite{James_Kwiat_Munro_White_2001},
\begin{equation}  \label{likelihood_to_Gauss}
  \sum_l f_l \ln p_l
  \rightarrow
  - \sum_l \frac{(f_l - p_l)^2}{2 \sigma_l^2}.
\end{equation}
The reconstruction based on this functional loses the essence of
the generalized measurement, nevertheless, it preserves all physical
properties of estimated quantum states.
The Gaussian limit of the likelihood method have been recently applied
to the reconstruction of polarization-entangled states of light
\cite{Nambu_et_al_2002,Nambu_et_al_2002_SPIE,Usami_et_al_2002}.
Unlike this, the approximate reconstruction of quantum processes
proposed in Ref.~\cite{Sacchi_2001_A} uses the exact likelihood
functional (\ref{process_likelihood}), however, it decreases the number
of the constraints  incorporated by the  Lagrange multipliers.
The $({\rm dim}{\cal H})^2$ necessary conditions that guarantee the correct
normalization of the estimated process are replaced by a single
condition, ${\rm Tr}[{S}] = {\rm dim}{\cal H}$. This is equivalent
to assuming that the Lagrange multiplier $\lambda$ is proportional to
identity operator.

\begin{figure}[ht]
 \centerline{\epsfig{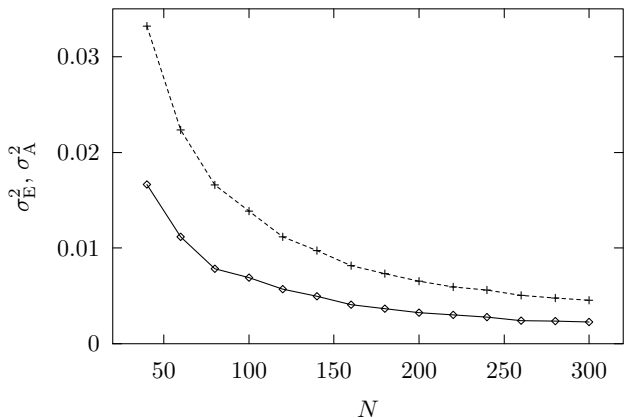}}
 \caption{The variance $\sigma_{\rm E}^2$ (diamond)
          of the exact maximum-likelihood reconstruction
          of a quantum process and the variance
          $\sigma_{\rm A}^2$ (plus) of the approximate one
          for various numbers $N$ of measurements.}
  \label{fig_comparison}
\end{figure}

In order to compare explicitly the exact maximum-likelihood estimation
of quantum process \cite{Fiurasek_Hradil_2001} with approximate method
presented in Refs.~\cite{Sacchi_2001_A,Sacchi_2001_B} we have carried out
extensive numerical simulations. Quantitative comparison
of the two approaches was based on the variances of estimates
${S}_{\rm E}$ (exact) and ${S}_{\rm A}$ (approximate),
\begin{equation}  \label{variances}
  \begin{array}{c}
    \sigma_{\rm E}^2 =
    \left\langle
      {\rm Tr}[({S}_{\rm E} - {S}_{\rm true})^2]
    \right\rangle_{\rm ens},
    \\
    \sigma_{\rm A}^2 =
    \left\langle 
      {\rm Tr}[({S}_{\rm A} - {S}_{\rm true})^2]
    \right\rangle_{\rm ens},
  \end{array}
\end{equation}
where $\langle\ldots\rangle_{\rm ens}$ denotes averaging
over an ensemble of all possible experimental data and
${S}_{\rm true}$ denotes the true CP map. For a given fixed
CP map, input states, and output measurements, we have repeated
$1000$ times a simulation of the measurements and reconstruction
of the CP maps ${S}_{\rm E}$  and ${S}_{\rm A}$. Subsequently
we have calculated variances (\ref{variances}) as statistical
averages over the acquired ensemble.
We have found that the exact maximum-likelihood estimation
yields in all cases much lower variance than approximate
approach. This is a direct consequence of the fact that the
exact treatment takes into account all constraints imposed
by quantum mechanical laws on the estimated operator ${S}$.
A typical example is shown in Fig.~\ref{fig_comparison}.
In this case, the quantum process is a unitary transformation
(\ref{pi4_rotation_process}) of a single qubit. Six different
input states are considered---eigenstates of three Pauli
matrices $\sigma_x$, $\sigma_y$, and $\sigma_z$. $3N$ copies
of each input state are used. On each corresponding output
state, a spin projection along axes $x$, $y$ and $z$ is measured
$N$ times. As can be seen in Fig.~\ref{fig_comparison}, the
variance $\sigma_{\rm E}^2$ is approximately twice smaller than
variance $\sigma_{\rm A}^2$, which is a significant difference.
In fact, for CP maps which do not represent unitary
transformations, such as Pauli damping channel, the difference
may be even stronger.

\section{Conclusions}

The unified approach to inference of quantum states and quantum
processes from experimental noisy data has been presented. The
proposed technique based on the maximum-likelihood principle
preserves all properties of the states and the processes imposed
by quantum mechanics. This method is very versatile and
can handle data from many different experimental configurations
such as the probing of quantum processes with entangled states
or a simultaneous reconstruction of an unknown process and
unknown states that are used to probe this process.
The extremal equations (\ref{state_extremal_sym_eq}),
(\ref{process_extremal_sym_eq})--%
(\ref{process_extremal_sym_eq_lambda}), and
(\ref{simultaneous_extremal_sym_eq__states})--%
(\ref{simultaneous_extremal_sym_eq_lambda})
for the most likely quantum state and process can be very
efficiently solved numerically by means of repeated iterations.
The exact maximum likelihood estimation of quantum objects
has been compared with the approximate methods. The
approximate ones yield estimates whose variance is typically
substantially larger than in the case of the exact approach.
This comparison clearly illustrates the importance of keeping
all the constraints imposed by quantum theory. Loosely speaking
there is always a choice---either to acquire less portion of
the data and then to adopt more sophisticated algorithm for
its evaluation or vice versa.
The efficient and precise reconstruction technique discussed
in the present paper can find applications in design and
evaluation of quantum-information devices and contemporary
quantum experiments.

\begin{acknowledgments}
  This work was supported by Grant LN00A015 and
  by Research Project CEZ: J14/98: 153100009
  ``Wave and particle optics''
  of the Czech Ministry of Education.
\end{acknowledgments}


\begin{thebibliography}{50}
\expandafter\ifx\csname natexlab\endcsname\relax\def\natexlab#1{#1}\fi
\expandafter\ifx\csname bibnamefont\endcsname\relax
  \def\bibnamefont#1{#1}\fi
\expandafter\ifx\csname bibfnamefont\endcsname\relax
  \def\bibfnamefont#1{#1}\fi
\expandafter\ifx\csname citenamefont\endcsname\relax
  \def\citenamefont#1{#1}\fi
\expandafter\ifx\csname url\endcsname\relax
  \def\url#1{\texttt{#1}}\fi
\expandafter\ifx\csname urlprefix\endcsname\relax\def\urlprefix{URL }\fi
\providecommand{\bibinfo}[2]{#2}
\providecommand{\eprint}[2][]{\url{#2}}

\bibitem[{\citenamefont{Vogel and Risken}(1989)}]{Vogel_Risken_1989}
\bibinfo{author}{\bibfnamefont{K.}~\bibnamefont{Vogel}} \bibnamefont{and}
  \bibinfo{author}{\bibfnamefont{H.}~\bibnamefont{Risken}},
  \bibinfo{journal}{Phys. Rev. A} \textbf{\bibinfo{volume}{40}},
  \bibinfo{pages}{2847} (\bibinfo{year}{1989}).

\bibitem[{\citenamefont{Smithey et~al.}(1993)\citenamefont{Smithey, Beck,
  Raymer, and Faridani}}]{Smithey_Beck_Raymer_Faridani_1993}
\bibinfo{author}{\bibfnamefont{D.~T.} \bibnamefont{Smithey}},
  \bibinfo{author}{\bibfnamefont{M.}~\bibnamefont{Beck}},
  \bibinfo{author}{\bibfnamefont{M.~G.} \bibnamefont{Raymer}},
  \bibnamefont{and} \bibinfo{author}{\bibfnamefont{A.}~\bibnamefont{Faridani}},
  \bibinfo{journal}{Phys. Rev. Lett.} \textbf{\bibinfo{volume}{70}},
  \bibinfo{pages}{1244} (\bibinfo{year}{1993}).

\bibitem[{\citenamefont{D'Ariano et~al.}(1995)\citenamefont{D'Ariano,
  Leonhardt, and Paul}}]{D'Ariano_Leonhardt_Paul_1995}
\bibinfo{author}{\bibfnamefont{G.~M.} \bibnamefont{D'Ariano}},
  \bibinfo{author}{\bibfnamefont{U.}~\bibnamefont{Leonhardt}},
  \bibnamefont{and} \bibinfo{author}{\bibfnamefont{H.}~\bibnamefont{Paul}},
  \bibinfo{journal}{Phys. Rev. A} \textbf{\bibinfo{volume}{52}},
  \bibinfo{pages}{R1801} (\bibinfo{year}{1995}).

\bibitem[{\citenamefont{Schiller et~al.}(1996)\citenamefont{Schiller,
  Breitenbach, Pereira, Muller, and Mlynek}}]{Schiller_Breitenbach_et_al_1996}
\bibinfo{author}{\bibfnamefont{S.}~\bibnamefont{Schiller}},
  \bibinfo{author}{\bibfnamefont{G.}~\bibnamefont{Breitenbach}},
  \bibinfo{author}{\bibfnamefont{S.~F.} \bibnamefont{Pereira}},
  \bibinfo{author}{\bibfnamefont{T.}~\bibnamefont{Muller}}, \bibnamefont{and}
  \bibinfo{author}{\bibfnamefont{J.}~\bibnamefont{Mlynek}},
  \bibinfo{journal}{Phys. Rev. Lett.} \textbf{\bibinfo{volume}{77}},
  \bibinfo{pages}{2933} (\bibinfo{year}{1996}).

\bibitem[{\citenamefont{Breitenbach et~al.}(1997)\citenamefont{Breitenbach,
  Schiller, and Mlynek}}]{Breitenbach_Schiller_Mlynek_1997}
\bibinfo{author}{\bibfnamefont{G.}~\bibnamefont{Breitenbach}},
  \bibinfo{author}{\bibfnamefont{S.}~\bibnamefont{Schiller}}, \bibnamefont{and}
  \bibinfo{author}{\bibfnamefont{J.}~\bibnamefont{Mlynek}},
  \bibinfo{journal}{Nature} \textbf{\bibinfo{volume}{387}},
  \bibinfo{pages}{471} (\bibinfo{year}{1997}).

\bibitem[{\citenamefont{Raymer et~al.}(1994)\citenamefont{Raymer, Beck, and
  McAlister}}]{Raymer_Beck_McAlister_1994}
\bibinfo{author}{\bibfnamefont{M.~G.} \bibnamefont{Raymer}},
  \bibinfo{author}{\bibfnamefont{M.}~\bibnamefont{Beck}}, \bibnamefont{and}
  \bibinfo{author}{\bibfnamefont{D.~F.} \bibnamefont{McAlister}},
  \bibinfo{journal}{Phys. Rev. Lett.} \textbf{\bibinfo{volume}{72}},
  \bibinfo{pages}{1137} (\bibinfo{year}{1994}).

\bibitem[{\citenamefont{Jones}(1994)}]{Jones_1994}
\bibinfo{author}{\bibfnamefont{K.~R.~W.} \bibnamefont{Jones}},
  \bibinfo{journal}{Phys. Rev. A} \textbf{\bibinfo{volume}{50}},
  \bibinfo{pages}{3682} (\bibinfo{year}{1994}).

\bibitem[{\citenamefont{D'Ariano et~al.}(1994)\citenamefont{D'Ariano,
  Macchiavello, and Paris}}]{D'Ariano_Macchiavello_Paris_1994}
\bibinfo{author}{\bibfnamefont{G.~M.} \bibnamefont{D'Ariano}},
  \bibinfo{author}{\bibfnamefont{C.}~\bibnamefont{Macchiavello}},
  \bibnamefont{and} \bibinfo{author}{\bibfnamefont{M.~G.~A.}
  \bibnamefont{Paris}}, \bibinfo{journal}{Phys. Rev. A}
  \textbf{\bibinfo{volume}{50}}, \bibinfo{pages}{4298} (\bibinfo{year}{1994}).

\bibitem[{\citenamefont{Dunn et~al.}(1995)\citenamefont{Dunn, Walmsley, and
  Mukamel}}]{Dunn_Walmsley_Mukamel_1995}
\bibinfo{author}{\bibfnamefont{T.~J.} \bibnamefont{Dunn}},
  \bibinfo{author}{\bibfnamefont{I.~A.} \bibnamefont{Walmsley}},
  \bibnamefont{and} \bibinfo{author}{\bibfnamefont{S.}~\bibnamefont{Mukamel}},
  \bibinfo{journal}{Phys. Rev. Lett.} \textbf{\bibinfo{volume}{74}},
  \bibinfo{pages}{884} (\bibinfo{year}{1995}).

\bibitem[{\citenamefont{Leonhardt et~al.}(1995)\citenamefont{Leonhardt, Paul,
  and D'Ariano}}]{Leonhardt_Paul_D'Ariano_1995}
\bibinfo{author}{\bibfnamefont{U.}~\bibnamefont{Leonhardt}},
  \bibinfo{author}{\bibfnamefont{H.}~\bibnamefont{Paul}}, \bibnamefont{and}
  \bibinfo{author}{\bibfnamefont{G.~M.} \bibnamefont{D'Ariano}},
  \bibinfo{journal}{Phys. Rev. A} \textbf{\bibinfo{volume}{52}},
  \bibinfo{pages}{4899} (\bibinfo{year}{1995}).

\bibitem[{\citenamefont{Leonhardt and Raymer}(1996)}]{Leonhardt_Raymer_1996}
\bibinfo{author}{\bibfnamefont{U.}~\bibnamefont{Leonhardt}} \bibnamefont{and}
  \bibinfo{author}{\bibfnamefont{M.~G.} \bibnamefont{Raymer}},
  \bibinfo{journal}{Phys. Rev. Lett.} \textbf{\bibinfo{volume}{76}},
  \bibinfo{pages}{1985} (\bibinfo{year}{1996}).

\bibitem[{\citenamefont{Banaszek and
  Wodkiewicz}(1996)}]{Banaszek_Wodkiewicz_1996}
\bibinfo{author}{\bibfnamefont{K.}~\bibnamefont{Banaszek}} \bibnamefont{and}
  \bibinfo{author}{\bibfnamefont{K.}~\bibnamefont{Wodkiewicz}},
  \bibinfo{journal}{Phys. Rev. Lett.} \textbf{\bibinfo{volume}{76}},
  \bibinfo{pages}{4344} (\bibinfo{year}{1996}).

\bibitem[{\citenamefont{Leibfried et~al.}(1996)\citenamefont{Leibfried,
  Meekhof, King, Monroe, Itano, and Wineland}}]{Leibfried_et_al_1996}
\bibinfo{author}{\bibfnamefont{D.}~\bibnamefont{Leibfried}},
  \bibinfo{author}{\bibfnamefont{D.~M.} \bibnamefont{Meekhof}},
  \bibinfo{author}{\bibfnamefont{B.~E.} \bibnamefont{King}},
  \bibinfo{author}{\bibfnamefont{C.}~\bibnamefont{Monroe}},
  \bibinfo{author}{\bibfnamefont{W.~M.} \bibnamefont{Itano}}, \bibnamefont{and}
  \bibinfo{author}{\bibfnamefont{D.~J.} \bibnamefont{Wineland}},
  \bibinfo{journal}{Phys. Rev. Lett.} \textbf{\bibinfo{volume}{77}},
  \bibinfo{pages}{4281} (\bibinfo{year}{1996}).

\bibitem[{\citenamefont{Kurtsiefer et~al.}(1997)\citenamefont{Kurtsiefer, Pfau,
  and Mlynek}}]{Kurtsiefer_Pfau_Mlynek_1997}
\bibinfo{author}{\bibfnamefont{C.}~\bibnamefont{Kurtsiefer}},
  \bibinfo{author}{\bibfnamefont{T.}~\bibnamefont{Pfau}}, \bibnamefont{and}
  \bibinfo{author}{\bibfnamefont{J.}~\bibnamefont{Mlynek}},
  \bibinfo{journal}{Nature} \textbf{\bibinfo{volume}{386}},
  \bibinfo{pages}{150} (\bibinfo{year}{1997}).

\bibitem[{\citenamefont{Lutterbach and
  Davidovich}(1997)}]{Lutterbach_Davidovich_1997}
\bibinfo{author}{\bibfnamefont{L.~G.} \bibnamefont{Lutterbach}}
  \bibnamefont{and}
  \bibinfo{author}{\bibfnamefont{L.}~\bibnamefont{Davidovich}},
  \bibinfo{journal}{Phys. Rev. Lett.} \textbf{\bibinfo{volume}{78}},
  \bibinfo{pages}{2547} (\bibinfo{year}{1997}).

\bibitem[{\citenamefont{Opatrn{\'{y}} and Welsch}(1997)}]{Opatrny_Welsch_1997}
\bibinfo{author}{\bibfnamefont{T.}~\bibnamefont{Opatrn{\'{y}}}}
  \bibnamefont{and} \bibinfo{author}{\bibfnamefont{D.~G.}
  \bibnamefont{Welsch}}, \bibinfo{journal}{Phys. Rev. A}
  \textbf{\bibinfo{volume}{55}}, \bibinfo{pages}{1462} (\bibinfo{year}{1997}).

\bibitem[{\citenamefont{Opatrn{\'{y}} et~al.}(1997)\citenamefont{Opatrn{\'{y}},
  Welsch, and Vogel}}]{Opatrny_Welsch_Vogel_1997}
\bibinfo{author}{\bibfnamefont{T.}~\bibnamefont{Opatrn{\'{y}}}},
  \bibinfo{author}{\bibfnamefont{D.~G.} \bibnamefont{Welsch}},
  \bibnamefont{and} \bibinfo{author}{\bibfnamefont{W.}~\bibnamefont{Vogel}},
  \bibinfo{journal}{Phys. Rev. A} \textbf{\bibinfo{volume}{56}},
  \bibinfo{pages}{1788} (\bibinfo{year}{1997}).

\bibitem[{\citenamefont{Bodendorf et~al.}(1998)\citenamefont{Bodendorf,
  Antesberger, Kim, and Walther}}]{Bodendorf_et_al_1998}
\bibinfo{author}{\bibfnamefont{C.~T.} \bibnamefont{Bodendorf}},
  \bibinfo{author}{\bibfnamefont{G.}~\bibnamefont{Antesberger}},
  \bibinfo{author}{\bibfnamefont{M.~S.} \bibnamefont{Kim}}, \bibnamefont{and}
  \bibinfo{author}{\bibfnamefont{H.}~\bibnamefont{Walther}},
  \bibinfo{journal}{Phys. Rev. A} \textbf{\bibinfo{volume}{57}},
  \bibinfo{pages}{1371} (\bibinfo{year}{1998}).

\bibitem[{\citenamefont{Lvovsky et~al.}(2001)\citenamefont{Lvovsky, Hansen,
  Aichele, Benson, Mlynek, and Schiller}}]{Lvovsky_Hansen_et_al_2001}
\bibinfo{author}{\bibfnamefont{A.~I.} \bibnamefont{Lvovsky}},
  \bibinfo{author}{\bibfnamefont{H.}~\bibnamefont{Hansen}},
  \bibinfo{author}{\bibfnamefont{T.}~\bibnamefont{Aichele}},
  \bibinfo{author}{\bibfnamefont{O.}~\bibnamefont{Benson}},
  \bibinfo{author}{\bibfnamefont{J.}~\bibnamefont{Mlynek}}, \bibnamefont{and}
  \bibinfo{author}{\bibfnamefont{S.}~\bibnamefont{Schiller}},
  \bibinfo{journal}{Phys. Rev. Lett.} \textbf{\bibinfo{volume}{87}},
  \bibinfo{pages}{050402} (\bibinfo{year}{2001}).

\bibitem[{\citenamefont{Lvovsky and Babichev}(2001)}]{Lvovsky_Babichev_2002}
\bibinfo{author}{\bibfnamefont{A.~I.} \bibnamefont{Lvovsky}} \bibnamefont{and}
  \bibinfo{author}{\bibfnamefont{S.~A.} \bibnamefont{Babichev}},
  \bibinfo{journal}{Phys. Rev. A} \textbf{\bibinfo{volume}{66}},
  \bibinfo{pages}{011801(R)} (\bibinfo{year}{2001}).

\bibitem[{\citenamefont{Opatrn{\'{y}} et~al.}(2002)\citenamefont{Opatrn{\'{y}},
  Korolkova, and Leuchs}}]{Opatrny_Korolkova_Leuchs_2002}
\bibinfo{author}{\bibfnamefont{T.}~\bibnamefont{Opatrn{\'{y}}}},
  \bibinfo{author}{\bibfnamefont{N.}~\bibnamefont{Korolkova}},
  \bibnamefont{and} \bibinfo{author}{\bibfnamefont{G.}~\bibnamefont{Leuchs}}
  (\bibinfo{year}{2002}), \bibinfo{note}{arXiv:quant-ph/0204131}.

\bibitem[{\citenamefont{Burlakov et~al.}(2002)\citenamefont{Burlakov,
  Krivitskiy, Kulik, Maslennikov, and Chekhova}}]{Burlakov_et_al_2002}
\bibinfo{author}{\bibfnamefont{A.~V.} \bibnamefont{Burlakov}},
  \bibinfo{author}{\bibfnamefont{L.~A.} \bibnamefont{Krivitskiy}},
  \bibinfo{author}{\bibfnamefont{S.~P.} \bibnamefont{Kulik}},
  \bibinfo{author}{\bibfnamefont{G.~A.} \bibnamefont{Maslennikov}},
  \bibnamefont{and} \bibinfo{author}{\bibfnamefont{M.~V.}
  \bibnamefont{Chekhova}} (\bibinfo{year}{2002}),
  \bibinfo{note}{arXiv:quant-ph/0207096}.

\bibitem[{\citenamefont{Hradil}(1997)}]{Hradil_1997}
\bibinfo{author}{\bibfnamefont{Z.}~\bibnamefont{Hradil}},
  \bibinfo{journal}{Phys. Rev. A} \textbf{\bibinfo{volume}{55}},
  \bibinfo{pages}{R1561} (\bibinfo{year}{1997}).

\bibitem[{\citenamefont{Hradil et~al.}(1999)\citenamefont{Hradil, Summhammer,
  and Rauch}}]{Hradil_Summhammer_Rauch_1999}
\bibinfo{author}{\bibfnamefont{Z.}~\bibnamefont{Hradil}},
  \bibinfo{author}{\bibfnamefont{J.}~\bibnamefont{Summhammer}},
  \bibnamefont{and} \bibinfo{author}{\bibfnamefont{H.}~\bibnamefont{Rauch}},
  \bibinfo{journal}{Phys. Lett. A} \textbf{\bibinfo{volume}{261}},
  \bibinfo{pages}{20} (\bibinfo{year}{1999}).

\bibitem[{\citenamefont{Hradil et~al.}(2000)\citenamefont{Hradil, Summhammer,
  Badurek, and Rauch}}]{Hradil_Summhammer_Badurek_Rauch_2000}
\bibinfo{author}{\bibfnamefont{Z.}~\bibnamefont{Hradil}},
  \bibinfo{author}{\bibfnamefont{J.}~\bibnamefont{Summhammer}},
  \bibinfo{author}{\bibfnamefont{G.}~\bibnamefont{Badurek}}, \bibnamefont{and}
  \bibinfo{author}{\bibfnamefont{H.}~\bibnamefont{Rauch}},
  \bibinfo{journal}{Phys. Rev A} \textbf{\bibinfo{volume}{62}},
  \bibinfo{pages}{014101} (\bibinfo{year}{2000}).

\bibitem[{\citenamefont{Hradil and Summhammer}(2000)}]{Hradil_Summhammer_2000}
\bibinfo{author}{\bibfnamefont{Z.}~\bibnamefont{Hradil}} \bibnamefont{and}
  \bibinfo{author}{\bibfnamefont{J.}~\bibnamefont{Summhammer}},
  \bibinfo{journal}{J. Phys. A: Math. Gen.} \textbf{\bibinfo{volume}{33}},
  \bibinfo{pages}{7607} (\bibinfo{year}{2000}).

\bibitem[{\citenamefont{Banaszek et~al.}(2000)\citenamefont{Banaszek, D'Ariano,
  Paris, and Sacchi}}]{Banaszek_D'Ariano_Paris_Sacchi_2000}
\bibinfo{author}{\bibfnamefont{K.}~\bibnamefont{Banaszek}},
  \bibinfo{author}{\bibfnamefont{G.~M.} \bibnamefont{D'Ariano}},
  \bibinfo{author}{\bibfnamefont{M.~G.~A.} \bibnamefont{Paris}},
  \bibnamefont{and} \bibinfo{author}{\bibfnamefont{M.~F.}
  \bibnamefont{Sacchi}}, \bibinfo{journal}{Phys. Rev. A}
  \textbf{\bibinfo{volume}{61}}, \bibinfo{pages}{10304(R)}
  (\bibinfo{year}{2000}).

\bibitem[{\citenamefont{{\v{R}}eh{\'{a}}{\v{c}}ek
  et~al.}(2001)\citenamefont{{\v{R}}eh{\'{a}}{\v{c}}ek, Hradil, and
  Je{\v{z}}ek}}]{Rehacek_Hradil_Jezek_2001}
\bibinfo{author}{\bibfnamefont{J.}~\bibnamefont{{\v{R}}eh{\'{a}}{\v{c}}ek}},
  \bibinfo{author}{\bibfnamefont{Z.}~\bibnamefont{Hradil}}, \bibnamefont{and}
  \bibinfo{author}{\bibfnamefont{M.}~\bibnamefont{Je{\v{z}}ek}},
  \bibinfo{journal}{Phys. Rev. A} \textbf{\bibinfo{volume}{63}},
  \bibinfo{pages}{040303(R)} (\bibinfo{year}{2001}).

\bibitem[{\citenamefont{{\v{R}}eh{\'{a}}{\v{c}}ek and
  Hradil}(2002)}]{Rehacek_Hradil_2002_INF}
\bibinfo{author}{\bibfnamefont{J.}~\bibnamefont{{\v{R}}eh{\'{a}}{\v{c}}ek}}
  \bibnamefont{and} \bibinfo{author}{\bibfnamefont{Z.}~\bibnamefont{Hradil}},
  \bibinfo{journal}{Phys. Rev. Lett.} \textbf{\bibinfo{volume}{88}},
  \bibinfo{pages}{130401} (\bibinfo{year}{2002}).

\bibitem[{\citenamefont{Poyatos et~al.}(1997)\citenamefont{Poyatos, Cirac, and
  Zoller}}]{Poyatos_Cirac_Zoller_1997}
\bibinfo{author}{\bibfnamefont{J.~F.} \bibnamefont{Poyatos}},
  \bibinfo{author}{\bibfnamefont{J.~I.} \bibnamefont{Cirac}}, \bibnamefont{and}
  \bibinfo{author}{\bibfnamefont{P.}~\bibnamefont{Zoller}},
  \bibinfo{journal}{Phys. Rev. Lett.} \textbf{\bibinfo{volume}{78}},
  \bibinfo{pages}{390} (\bibinfo{year}{1997}).

\bibitem[{\citenamefont{Chuang and Nielsen}(1997)}]{Chuang_Nielsen_1997}
\bibinfo{author}{\bibfnamefont{I.~L.} \bibnamefont{Chuang}} \bibnamefont{and}
  \bibinfo{author}{\bibfnamefont{M.~A.} \bibnamefont{Nielsen}},
  \bibinfo{journal}{J.~Mod Opt.} \textbf{\bibinfo{volume}{44}},
  \bibinfo{pages}{2455} (\bibinfo{year}{1997}).

\bibitem[{\citenamefont{D'Ariano and Maccone}(1998)}]{D'Ariano_Maccone_1998}
\bibinfo{author}{\bibfnamefont{G.~M.} \bibnamefont{D'Ariano}} \bibnamefont{and}
  \bibinfo{author}{\bibfnamefont{L.}~\bibnamefont{Maccone}},
  \bibinfo{journal}{Phys. Rev. Lett.} \textbf{\bibinfo{volume}{80}},
  \bibinfo{pages}{5465} (\bibinfo{year}{1998}).

\bibitem[{\citenamefont{Luis and
  S\'{a}nchez-Soto}(1999)}]{Luis_Sanchez-Soto_1999}
\bibinfo{author}{\bibfnamefont{A.}~\bibnamefont{Luis}} \bibnamefont{and}
  \bibinfo{author}{\bibfnamefont{L.~L.} \bibnamefont{S\'{a}nchez-Soto}},
  \bibinfo{journal}{Phys. Lett. A} \textbf{\bibinfo{volume}{261}},
  \bibinfo{pages}{12} (\bibinfo{year}{1999}).

\bibitem[{\citenamefont{Gutzeit et~al.}(2000)\citenamefont{Gutzeit,
  Wallentowitz, and Vogel}}]{Gutzeit_Wallentowitz_Vogel_2000}
\bibinfo{author}{\bibfnamefont{R.}~\bibnamefont{Gutzeit}},
  \bibinfo{author}{\bibfnamefont{S.}~\bibnamefont{Wallentowitz}},
  \bibnamefont{and} \bibinfo{author}{\bibfnamefont{W.}~\bibnamefont{Vogel}},
  \bibinfo{journal}{Phys. Rev. A} \textbf{\bibinfo{volume}{61}},
  \bibinfo{pages}{062105} (\bibinfo{year}{2000}).

\bibitem[{\citenamefont{Childs et~al.}(2001)\citenamefont{Childs, Chuang, and
  Leung}}]{Childs_Chuang_Leung_2001}
\bibinfo{author}{\bibfnamefont{A.~M.} \bibnamefont{Childs}},
  \bibinfo{author}{\bibfnamefont{I.~L.} \bibnamefont{Chuang}},
  \bibnamefont{and} \bibinfo{author}{\bibfnamefont{D.~W.} \bibnamefont{Leung}},
  \bibinfo{journal}{Phys. Rev. A} \textbf{\bibinfo{volume}{64}},
  \bibinfo{pages}{012314} (\bibinfo{year}{2001}).

\bibitem[{\citenamefont{Fischer et~al.}(2001)\citenamefont{Fischer, Mack,
  Cirone, and Freyberger}}]{Fischer_Mack_Cirone_Freyberger_2001}
\bibinfo{author}{\bibfnamefont{D.~G.} \bibnamefont{Fischer}},
  \bibinfo{author}{\bibfnamefont{H.}~\bibnamefont{Mack}},
  \bibinfo{author}{\bibfnamefont{M.~A.} \bibnamefont{Cirone}},
  \bibnamefont{and}
  \bibinfo{author}{\bibfnamefont{M.}~\bibnamefont{Freyberger}},
  \bibinfo{journal}{Phys. Rev. A} \textbf{\bibinfo{volume}{64}},
  \bibinfo{pages}{022309} (\bibinfo{year}{2001}).

\bibitem[{\citenamefont{D'Ariano and {Lo
  Presti}}(2001)}]{D'Ariano_LoPresti_2001}
\bibinfo{author}{\bibfnamefont{G.~M.} \bibnamefont{D'Ariano}} \bibnamefont{and}
  \bibinfo{author}{\bibfnamefont{P.}~\bibnamefont{{Lo Presti}}},
  \bibinfo{journal}{Phys. Rev. Lett.} \textbf{\bibinfo{volume}{86}},
  \bibinfo{pages}{4195} (\bibinfo{year}{2001}).

\bibitem[{\citenamefont{Fiur{\'{a}}{\v{s}}ek and
  Hradil}(2001)}]{Fiurasek_Hradil_2001}
\bibinfo{author}{\bibfnamefont{J.}~\bibnamefont{Fiur{\'{a}}{\v{s}}ek}}
  \bibnamefont{and} \bibinfo{author}{\bibfnamefont{Z.}~\bibnamefont{Hradil}},
  \bibinfo{journal}{Phys. Rev. A} \textbf{\bibinfo{volume}{63}},
  \bibinfo{pages}{020101(R)} (\bibinfo{year}{2001}).

\bibitem[{\citenamefont{James et~al.}(2001)\citenamefont{James, Kwiat, Munro,
  and White}}]{James_Kwiat_Munro_White_2001}
\bibinfo{author}{\bibfnamefont{D.~F.~V.} \bibnamefont{James}},
  \bibinfo{author}{\bibfnamefont{P.~G.} \bibnamefont{Kwiat}},
  \bibinfo{author}{\bibfnamefont{W.~J.} \bibnamefont{Munro}}, \bibnamefont{and}
  \bibinfo{author}{\bibfnamefont{A.~G.} \bibnamefont{White}},
  \bibinfo{journal}{Phys. Rev. A} \textbf{\bibinfo{volume}{64}},
  \bibinfo{pages}{052312} (\bibinfo{year}{2001}).

\bibitem[{\citenamefont{Sacchi}(2001{\natexlab{a}})}]{Sacchi_2001_A}
\bibinfo{author}{\bibfnamefont{M.~F.} \bibnamefont{Sacchi}},
  \bibinfo{journal}{Phys. Rev. A} \textbf{\bibinfo{volume}{63}},
  \bibinfo{pages}{054104} (\bibinfo{year}{2001}{\natexlab{a}}).

\bibitem[{\citenamefont{Helstrom}(1976)}]{Helstrom_QDET}
\bibinfo{author}{\bibfnamefont{C.~W.} \bibnamefont{Helstrom}},
  \emph{\bibinfo{title}{Quantum Detection and Estimation Theory}}
  (\bibinfo{publisher}{Academic Press}, \bibinfo{address}{New York},
  \bibinfo{year}{1976}).

\bibitem[{\citenamefont{Fiur{\'{a}}{\v{s}}ek}(2001)}]{Fiurasek_2001_POVM}
\bibinfo{author}{\bibfnamefont{J.}~\bibnamefont{Fiur{\'{a}}{\v{s}}ek}},
  \bibinfo{journal}{Phys. Rev. A} \textbf{\bibinfo{volume}{64}},
  \bibinfo{pages}{024102} (\bibinfo{year}{2001}).

\bibitem[{\citenamefont{Je{\v{z}}ek}(2002)}]{Jezek_2002}
\bibinfo{author}{\bibfnamefont{M.}~\bibnamefont{Je{\v{z}}ek}},
  \bibinfo{journal}{Phys. Lett. A} \textbf{\bibinfo{volume}{299}},
  \bibinfo{pages}{441} (\bibinfo{year}{2002}).

\bibitem[{\citenamefont{Sacchi}(2001{\natexlab{b}})}]{Sacchi_2001_B}
\bibinfo{author}{\bibfnamefont{M.~F.} \bibnamefont{Sacchi}},
  \bibinfo{journal}{Phys. Rev. A} \textbf{\bibinfo{volume}{64}},
  \bibinfo{pages}{022106} (\bibinfo{year}{2001}{\natexlab{b}}).

\bibitem[{\citenamefont{Hellwig and Kraus}(1970)}]{Hellwig_Kraus_1970}
\bibinfo{author}{\bibfnamefont{K.}~\bibnamefont{Hellwig}} \bibnamefont{and}
  \bibinfo{author}{\bibfnamefont{K.}~\bibnamefont{Kraus}},
  \bibinfo{journal}{Commun. Math. Phys.} \textbf{\bibinfo{volume}{16}},
  \bibinfo{pages}{142} (\bibinfo{year}{1970}).

\bibitem[{\citenamefont{Jamiolkowski}(1972)}]{Jamiolkowski_1972}
\bibinfo{author}{\bibfnamefont{A.}~\bibnamefont{Jamiolkowski}},
  \bibinfo{journal}{Rep. Math. Phys.} \textbf{\bibinfo{volume}{3}},
  \bibinfo{pages}{275} (\bibinfo{year}{1972}).

\bibitem[{\citenamefont{Schumacher}(1996)}]{Schumacher_1996}
\bibinfo{author}{\bibfnamefont{B.}~\bibnamefont{Schumacher}},
  \bibinfo{journal}{Phys. Rev. A} \textbf{\bibinfo{volume}{54}},
  \bibinfo{pages}{2614} (\bibinfo{year}{1996}).

\bibitem[{\citenamefont{Nambu et~al.}(2002{\natexlab{a}})\citenamefont{Nambu,
  Usami, Tsuda, Matsumoto, and Nakamura}}]{Nambu_et_al_2002}
\bibinfo{author}{\bibfnamefont{Y.}~\bibnamefont{Nambu}},
  \bibinfo{author}{\bibfnamefont{K.}~\bibnamefont{Usami}},
  \bibinfo{author}{\bibfnamefont{Y.}~\bibnamefont{Tsuda}},
  \bibinfo{author}{\bibfnamefont{K.}~\bibnamefont{Matsumoto}},
  \bibnamefont{and} \bibinfo{author}{\bibfnamefont{K.}~\bibnamefont{Nakamura}},
  \bibinfo{journal}{Phys. Rev. A} \textbf{\bibinfo{volume}{66}},
  \bibinfo{pages}{033816} (\bibinfo{year}{2002}{\natexlab{a}}).

\bibitem[{\citenamefont{Nambu et~al.}(2002{\natexlab{b}})\citenamefont{Nambu,
  Usami, Tomita, Ishizaka, Hiroshima, Tsuda, Matsumoto, and
  Nakamura}}]{Nambu_et_al_2002_SPIE}
\bibinfo{author}{\bibfnamefont{Y.}~\bibnamefont{Nambu}},
  \bibinfo{author}{\bibfnamefont{K.}~\bibnamefont{Usami}},
  \bibinfo{author}{\bibfnamefont{A.}~\bibnamefont{Tomita}},
  \bibinfo{author}{\bibfnamefont{S.}~\bibnamefont{Ishizaka}},
  \bibinfo{author}{\bibfnamefont{T.}~\bibnamefont{Hiroshima}},
  \bibinfo{author}{\bibfnamefont{Y.}~\bibnamefont{Tsuda}},
  \bibinfo{author}{\bibfnamefont{K.}~\bibnamefont{Matsumoto}},
  \bibnamefont{and} \bibinfo{author}{\bibfnamefont{K.}~\bibnamefont{Nakamura}},
  in \emph{\bibinfo{booktitle}{Quantum Optics in Computing and Communications,
  {\rm Proceedings of SPIE}}}, edited by
  \bibinfo{editor}{\bibfnamefont{S.}~\bibnamefont{Liu}},
  \bibinfo{editor}{\bibfnamefont{G.}~\bibnamefont{Guo}},
  \bibinfo{editor}{\bibfnamefont{H.-K.} \bibnamefont{Lo}}, \bibnamefont{and}
  \bibinfo{editor}{\bibfnamefont{N.}~\bibnamefont{Imoto}}
  (\bibinfo{year}{2002}{\natexlab{b}}), vol. \bibinfo{volume}{4917},
  p.~\bibinfo{pages}{13}.

\bibitem[{\citenamefont{Usami et~al.}(2002)\citenamefont{Usami, Nambu, Tsuda,
  Matsumoto, and Nakamura}}]{Usami_et_al_2002}
\bibinfo{author}{\bibfnamefont{K.}~\bibnamefont{Usami}},
  \bibinfo{author}{\bibfnamefont{Y.}~\bibnamefont{Nambu}},
  \bibinfo{author}{\bibfnamefont{Y.}~\bibnamefont{Tsuda}},
  \bibinfo{author}{\bibfnamefont{K.}~\bibnamefont{Matsumoto}},
  \bibnamefont{and} \bibinfo{author}{\bibfnamefont{K.}~\bibnamefont{Nakamura}}
  (\bibinfo{year}{2002}), \bibinfo{note}{arXiv:quant-ph/0209074}.

\end{thebibliography}

\end{document}